\documentclass[10pt, aps, prl, amsmath, amssymb, superscriptaddress, onecolumn]{revtex4-2}
\usepackage{xcolor,soul,graphicx}
\usepackage[colorlinks=true,citecolor=blue,urlcolor=blue]{hyperref}

\newcommand\R{\mathrm{R}}
\newcommand\Prm{\mathrm{P}}
\newcommand\Srm{\mathrm{S}}

\begin{document}

\title{Active waves from non-reciprocity and cytoplasmic exchange}

\author{Jason R. Picardo}
\email{picardo@iitb.ac.in}
\affiliation{Department of Chemical Engineering,
Indian Institute of Technology Bombay, Powai, Mumbai, India 400076}

\author{V. Jemseena}
\affiliation{International Centre for Theoretical Sciences, Tata Institute of Fundamental Research, \\ 
Survey 151, Shivakote Village, Hesaraghatta Hobli, 
Bengaluru North, India 560089.}

\author{K. Vijay Kumar}
\email{vijaykumar@icts.res.in}
\affiliation{International Centre for Theoretical Sciences, Tata Institute of Fundamental Research, \\ 
Survey 151, Shivakote Village, Hesaraghatta Hobli, 
Bengaluru North, India 560089.}


\begin{abstract}
Pattern formation in active biological matter typically arises from the feedback between chemical concentration fields and mechanical stresses. The actomyosin cortex of cells is an archetypal example of an active thin film that displays such patterns. Here, we show how pulsatory patterns emerge in a minimal model of the actomyosin cortex with a single stress-regulating chemical species that exchanges material with the cytoplasm via a linear turnover reaction. Deriving a low-dimensional amplitude-phase model, valid for a one-dimensional periodic domain and 
a spherical surface, we show that nonlinear waves arise from a secondary parity-breaking bifurcation that originates from the nonreciprocal interaction between spatial modes of the concentration field.  Numerical analysis confirms these analytical predictions, and also reveals analogous pulsatory patterns on impermeable domains. Our study provides a generic route to the emergence of nonreciprocity-driven pulsatory patterns that can be controlled by both the strength of activity and the turnover rate.
\end{abstract}

\maketitle

\textbf{Introduction.} Pattern formation is ubiquitous in living matter \cite{Cross2009,MurrayII2003,Meinhardt, Pismen2023} and has been studied extensively in the framework of reaction-diffusion equations \cite{EpsteinPojman1998,kondoReactiondiffusionModelFramework2010}. For example, Turing patterns \cite{turingChemicalBasisMorphogenesis1952}, implicated in many developmental processes \cite{Sharpe,Harrison}, emerge when a fast diffusing autocatalytic activator interacts with a slow diffusing inhibitor. However, many significant biological patterns at the cell and tissue levels involve a strong coupling between biochemical signaling and mechanical forces---morphogenesis being a prime example \cite{DarcyThompson,howardTuringsNextSteps2011,grossHowActiveMechanics2017}. The fundamental agency responsible for the generation of mechanical forces in living matter is the Adenosine Triphosphate (ATP) consuming activity of molecular motors in the eukaryotic cytoskeleton \cite{Howard2001, Kolomeisky2015}. At large scales, these nonequilibrium processes manifest as active stresses \cite{marchettiHydrodynamicsSoftActive2013,julicherHydrodynamicTheoryActive2018}. Gradients in these active stresses drive hydrodynamic flows or deformations and can lead to the emergence of  mechanochemical patterns \cite{boisPatternFormationActive2011,kumarPulsatoryPatternsActive2014,mietkeMinimalModelCellular2019}. For example, flows in the actomyosin cortex are known to be involved in the establishment of developmental patterns \cite{Mayer2010,Goehring2011,Gross2019,baillesGeneticInductionMechanochemical2019,Collinet2023.05.02.539070}.

The actomyosin cortex -- a thin film ($\sim 0.5 \mu\mathrm{m}$) beneath the cell membrane consisting of actin filaments, myosin molecular motors, and cross-linking proteins -- is a dynamic material that undergoes constant turnover (on timescales $\sim 30\mathrm{s}$) due to the polymerization and depolymerization of its constituents \cite{fritzscheActinKineticsShapes2016, chughActinCortexGlance2018, kumarActomyosinCortexCells2021}.  On timescales longer than the viscoelastic relation time ($\sim 5 \mathrm{s}$) \cite{sahaDeterminingPhysicalProperties2016}, the actomyosin cortex flows like a fluid  in response to gradients in active contractility, regulated by myosin concentration. The feedback between such active flows, diffusive transport, and chemical reactions leads to the emergence of patterns. 

Actomyosin patterns are typically pulsatory and their excitable character underlies many cellular processes \cite{martinPulsedContractionsActinmyosin2009,rauziPlanarPolarizedActomyosin2010,nishikawaControllingContractileInstabilities2017,yangRhythmicityWavesCortex2018,chanetCollectiveCellSorting2020,michaudCorticalExcitabilityCell2021}. Pulsatile active patterns have been observed in theoretical models involving multiple active species \cite{kumarPulsatoryPatternsActive2014,barberiLocalizedSpatiotemporalDynamics2024}, orientational order parameters \cite{husainEmergentStructuresActive2017}, and viscoelasticity \cite{dierkesSpontaneousOscillationsElastic2014,alonsoMechanochemicalPatternFormation2017,banerjeeActomyosinPulsationFlows2017,Mackay2024.10.04.616649}. Despite their differences, these models share a common route to oscillatory patterns-- namely a Hopf bifurcation of the homogeneous state. 

In this study, we show how pulsatory patterns arise 
in an active film with an isotropic stress that is regulated by a single chemical species, which undergoes a linear turnover reaction. Since the concentration of the active species is the only dynamic field, the primary bifurcation of the homogeneous state is non-oscillatory and leads to stationary patterns. However, at sufficiently high activity, i.e. in the strongly nonlinear regime, a parity-breaking secondary bifurcation can result in the emergence of travelling waves. Considering a one-dimensional periodic domain, as well as the surface of a sphere, we use a Galerkin reduced model in conjunction with numerical analysis to show that these waves arise due to nonlinear and non-reciprocal interactions between different spatial modes of the concentration field. Remarkably, the time-dependent patterns can be switched on and off by tuning the rate of the \textit{prima facie} stabilizing turnover reaction. Our work thus provides a generic mechanism for the generation and control of active pulsatory patterns in fluid films with a lone active species.

\begin{table*}
{
\centering
\begingroup
\renewcommand{\arraystretch}{1.5}
\setlength{\tabcolsep}{4pt}
\begin{tabular}{| c | c c c c c c c c c c c c |}
\hline
Coefficient & $m_1$ & $\mathcal{A}_1$ & $\mathcal{B}_1$ & $\mathcal{C}_1$ & $\mathcal{D}_1$ & $m_2$ & $\mathcal{A}_2$ & $\mathcal{B}_2$ & $\mathcal{C}_2$ & $\mathcal{D}_2$ & $\mathcal{A}_\varphi$ & $\mathcal{B}_\varphi$ 
\\\hline
Line & {\scriptsize $1$} & $\frac{\Srm-2}{\alpha_1}$ & $\frac{1}{2 \alpha_4}$ & $\frac{-\alpha_{-3}}{\alpha_1 \alpha_9}$ &  $\frac{-\alpha_{-2}\Srm+4 \alpha_1}{2 \alpha_1 \alpha_4}$ & {\scriptsize $4$} & $\frac{4(\Srm-2)}{\alpha_4}$ &  $\frac{4 \alpha_3}{\alpha_1 \alpha_9}$ &  $\frac{2}{\alpha_{16}}$ & $\frac{ \alpha_4 (\Srm-2) -2 \alpha_1}{\alpha_1 \alpha_4}$ &  $\frac{\alpha_4 \Srm -\alpha_{10}-3}{\alpha_1 \alpha_4}$ &  $2 \mathcal{D}_1$  
\\[.5em]
Sphere & {\scriptsize$2$} & $\frac{2(\Srm-2)}{\alpha_6}$ & $\frac{9}{10 \pi \alpha_{22}}$ & $\frac{3}{5 \pi \alpha_6}$ &  $\sqrt{\frac{6}{5\pi}}\frac{( -\alpha_{-2} \Srm+4\alpha_{10})}{\alpha_6 \alpha_{22}}$ & {\scriptsize $6$} &  $\frac{6(\Srm-2)}{\alpha_{22}}$ & { $3 \mathcal{C}_1$} &  {\scriptsize 0} &  $\sqrt{\frac{27}{10\pi}}\frac{(\alpha_{22}\Srm-4 \alpha_{14})}{\alpha_6 \alpha_{22}}$ & $\mathcal{D}_2$ &  { $2 \mathcal{D}_1$}
\\[.4em]
\hline
\end{tabular}
\endgroup
\caption{Coefficients of the reduced model on a periodic one-dimensional domain and on the surface of a sphere. Here $\alpha_j = 1 + j \, \mu^2$ is a linear function of $\mu^2$ with $j \in \mathbb{Z}$.}
\label{tab:coefficients}
}
\end{table*}

\textbf{Active fluid model.} 
In a one dimensional periodic domain of size $2\pi L$, the dynamics of the surface concentration $c(x,t)$ of a single regulator of active stress (e.g., a motor protein like myosin in the cell cortex) is governed by an advection-diffusion equation
\begin{equation}
\partial_t c = -\partial_x (vc) + D \partial_x^2 c - \kappa \, (c-c_0),
\label{eqn:conc}
\end{equation}
where $D$ is the diffusion coefficient, and the linear chemical reaction, with a base concentration $c_0$ and a rate $\kappa$, models the exchange between the cell surface and the cytoplasm. Neglecting inertial forces (appropriate for cellular scales), the hydrodynamic velocity $v$ is obtained from the force-balance equation $\partial_x \sigma = \gamma v$, where $\sigma$ is the total hydrodynamic stress and $\gamma$ is a friction coefficient that models the drag due to relative motion between the cortex and the cytoplasm and/or cell membrane. With a Newtonian constitutive relation for the passive stress, we get
\begin{equation}
\eta \partial_x^2 v - \gamma v = -\partial_x \sigma^a(c),
\label{eqn:mom}
\end{equation}
where $\eta$ is a viscosity coefficient. The active contractile stress $\sigma^a(c) = \zeta \Delta \mu \, f(c)$, where $\Delta \mu$ is the chemical potential difference of ATP hydrolysis, $\zeta$ is the strength of activity, and $f(c)=c/(c_s+c)$ with $c_s$ being a saturation concentration. On scaling lengths, times, and concentrations by $L$, $L^2/D$, and $c_0$, respectively, and transforming to non-dimensional variables, the model equations \eqref{eqn:conc} and \eqref{eqn:mom} are parametrized by: (1) the P{\'e}clet number $\Prm \equiv \zeta \Delta \mu/(\gamma D) $  measuring the strength of the active advection to diffusion, (2) the Dahmk{\"o}hler number $\R \equiv \kappa L^2/D$ comparing the turnover rate with the diffusion timescale across the domain, (3) $\mu \equiv \sqrt{\eta/(\gamma L^2)}$, and (4) $\Srm \equiv c_{s}/c_0$.

\textbf{Linear and weakly-nonlinear stability.} The model in \eqref{eqn:conc} and \eqref{eqn:mom} leads to the emergence of stationary patterns for sufficiently large $\Prm$, even with $\R=0$  \cite{boisPatternFormationActive2011}. When $\R \neq 0$, though, the model also exhibits nonlinear time-dependent patterns \cite{mooreSelfOrganizingActomyosinPatterns2014}. This is remarkable for two reasons. First, an integration of \eqref{eqn:conc} over the spatial domain shows that the mean concentration $(2\pi)^{-1} \, \int_0^{2\pi} dx \, c(x,t)$ decays exponentially in time. As such, any time-dependent patterns should arise from the nonlinear coupling between the various spatial modes of the concentration. Second, the homogeneous state $c = 1$ cannot directly transition to time-dependent patterns. Indeed, a linear stability analysis of this state, using perturbations of the form $\exp (\lambda t + i k x)$, shows that the growth rate $\lambda$ corresponding to the wavenumber $k$ is
\begin{equation}
\lambda(k) = k^2 \left(\frac{\Prm \, \Srm}{(1 + \mu^2 k^2) \, (1+\Srm)^2} - 1 \right) - \R.
\label{eqn:disp}
\end{equation}
Clearly, $\lambda$ is real, and the homogeneous state undergoes a type-Is instability at a wavenumber $k_c$ beyond a critical P\'eclet  value $\Prm_c$ \cite{Cross2009}. Furthermore, for $\Prm \gtrapprox \Prm_c$, a weakly nonlinear analysis \cite{Cross2009,Roberts2015} shows that the amplitude $A$ of the fastest growing mode satisfies the real Ginzburg-Landau equation: 
\begin{equation}
\partial_t A = \Lambda A - \beta A^3 + \mathcal{D} \, \nabla^2 A ,\label{eqn:ampeqn}
\end{equation}
where $\Lambda=\lambda(k_c)$, and $\beta$ and $\mathcal{D}$ are positive constants, and $\beta$ increases with $\R$. Hence, close to the onset of the instability, the inhomogeneous patterned state is \emph{non-oscillatory} and stable to small perturbations. To summarize, both linear and weakly-nonlinear analysis of the homogeneous state predict a \textit{non-oscillatory} instability that is suppressed by the turnover rate $\R$.

\textbf{Nonlinear bifurcations.} To analytically investigate time-dependent patterns in the strongly nonlinear regime, we construct a Galerkin truncated model for the concentration field \citep{Balmforth2001}. In a one-dimensional periodic domain, we consider an $N$-mode Fourier expansion $c(x,t) = 1 + \sum_{n=1}^N \rho_n(t) \, \, \cos\left(n \left[x - \varphi_n(t) \right]\, \right)$ where the amplitudes $\rho_n$ and the phases $\varphi_n$ are real numbers. Since the $n=0$ mode decays exponentially ($\sim e^{-\R t}$), we have neglected its transient dynamics and used $\rho_0 = 1$. Substituting the above expansion in \eqref{eqn:conc}, and assuming $S \gg 1$, leads to a hierarchy of coupled equations for $\rho_n$ and $\varphi_n$ (see the supplementary information (SI) \citep{supplement}). A one-mode ($N=1$) truncation predicts only stationary patterns: the homogeneous state ($\rho_1=0$ and an arbitrary $\varphi_1$) undergoes a forward pitchfork bifurcation to a stationary patterned state ($\rho_1 \neq 0$ and an arbitrary $\varphi_1$) beyond a critical $\Prm$ \citep{supplement}.

\begin{figure*}
\includegraphics[width=\linewidth]{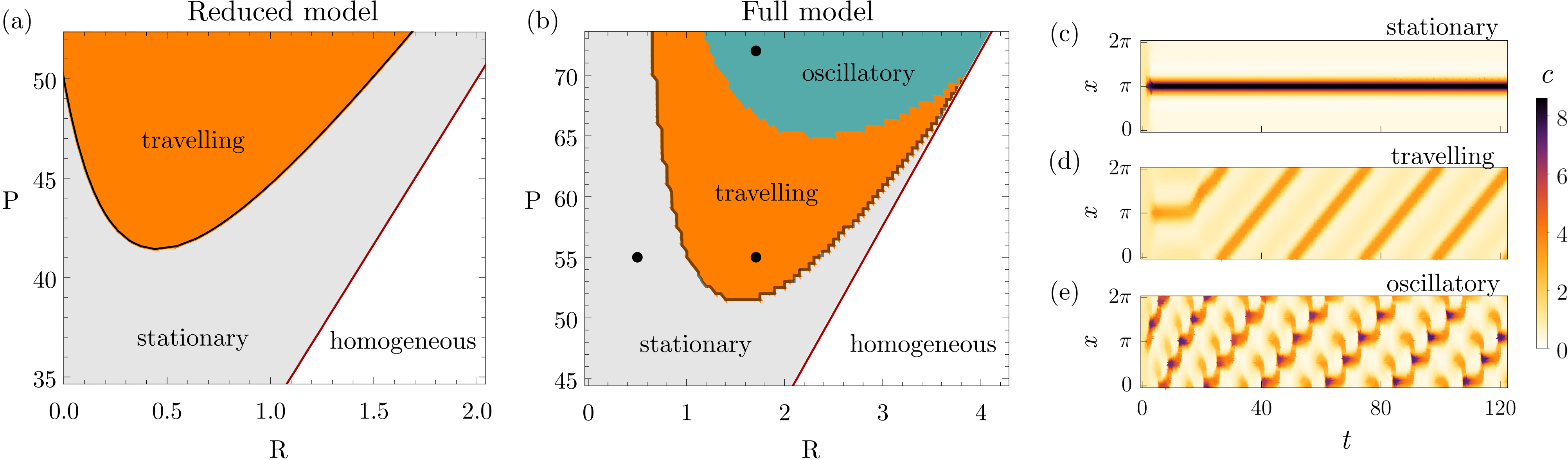}
\caption{Phase diagrams in the $\Prm-\R$ plane on a one-dimensional periodic domain, obtained from (a) the reduced model [\eqref{eq:rho1_pbc}-\eqref{eq:phi_pbc}], and (b) numerical solutions of the full model [\eqref{eqn:conc}-\eqref{eqn:mom}]. Representative kymographs of the concentration field $c(x,t)$ (in the full model) are presented in panels (c-e), along with movies in the SI \citep{supplement}. Here $\mu=1$ and $\Srm=5$.}
\label{fig:Kybif}
\end{figure*}

When $N=2$, we obtain
\begin{align}
\dot{\rho}_1 &= - (\R + m_1) \rho_1 
+ \frac{\Prm}{\Srm^2} \, \bigg[ 
\rho_1  
\left(\mathcal{A}_1  - \mathcal{B}_1 \rho_1^2 + \mathcal{C}_1 \rho_2^2 
\right)
- \mathcal{D}_1 \rho_1 \rho_2 \cos\varphi \bigg],
\label{eq:rho1_pbc}
\\
\dot{\rho}_2 &= - (\R + m_2) \rho_2 
+ \frac{\Prm}{\Srm^2}  \bigg[
\rho_2
\left(\mathcal{A}_2 - \mathcal{B}_2 \rho_1^2  - \mathcal{C}_2 \rho_2^2 
\right)
+ \mathcal{D}_2 \rho_1^2 \cos\varphi \bigg],
\label{eq:rho2_pbc}
\end{align}
where the phase-difference $\varphi = 2 (\varphi_2-\varphi_1)$ satisfies
\begin{align}
\dot{\varphi} &= -\frac{\Prm}{\Srm^2} \frac{(\mathcal{A}_{\varphi} \rho_1^2 -  \mathcal{B}_{\varphi} \rho_2^2)}{\rho_2} \, \sin \varphi.
\label{eq:phi_pbc}
\end{align}
Here, the coefficients $m_1$, $\mathcal{A}_1$ $\ldots$ are functions of $\mu$ and $\Srm$, and are listed in the first row of Tab.~\ref{tab:coefficients}.  The \emph{reduced model}, comprising \eqref{eq:rho1_pbc}, \eqref{eq:rho2_pbc}, and \eqref{eq:phi_pbc}, has three fixed points: (i) homogeneous states, $\rho_n = 0$ and arbitrary $\varphi$ with $\dot{\varphi}_n = 0$, (ii) stationary patterns, $\rho_n \neq 0$ and $\varphi=\{0,\pi\}$ with $\dot{\varphi}_n = 0$, and (iii) $\rho_n \neq 0$ and $0 < \varphi < \pi$ with $\dot{\varphi}_1=\dot{\varphi}_2=u$. Importantly, this third set corresponds to travelling waves, with a concentration profile $c(x,t) = 1 + \rho_1 \cos\left[ x - u t \right] + \rho_2 \cos\left[2(x - u t) + \varphi\right]$, that translate linearly in time with constant speed $u$ (assuming $\varphi_1(0)=0$). A linear stability analysis shows that these fixed points have distinct non-overlapping regions of stability \citep{supplement} and leads to the phase-diagram shown in Fig. \ref{fig:Kybif}(a). We see that in addition to the primary bifurcation of the homogeneous state, the stationary pattern undergoes a secondary bifurcation to travelling waves. At moderately high activity ($\Prm \sim 45$), increasing the turnover rate $\R$ causes a transition from stationary patterns to traveling waves and then again to stationary patterns, before eventually stabilizing the homogeneous state.

Motivated by these analytical predictions, we numerically solve the \emph{full model}, i.e., equations \eqref{eqn:conc} and \eqref{eqn:mom}. These results are summarized in the phase diagram shown in Fig.~\ref{fig:Kybif}(b), which bears a strong resemblance to
that of the reduced model [Fig.~\ref{fig:Kybif}(a)]. The spatiotemporal dynamics of $c(x,t)$ in the various phases is illustrated in Figs.~\ref{fig:Kybif}(c-e) (see \cite{supplement} for movies). We observe the following points. First, the phase-boundary between travelling waves and stationary patterns asymptotes to the linear stability boundary of the homogeneous state. This feature, found in both the reduced model and the full model, is a result of the homogeneous state undergoing a type-Is instability (see Eq. \eqref{eqn:disp}). Second, at large $\Prm$, the full model exhibits a further bifurcation of travelling waves to oscillatory patterns without a fixed shape. It is possible that the full model exhibits more complex spatiotemporal dynamics at even higher $P$. Third, travelling waves do not occur when $\R=0$ in the full model [Fig.~\ref{fig:Kybif}(b)]. Indeed, in the limit $\mu \ll 1$, we can analytically demonstrate that the full model does not admit travelling wave solutions in the absence of turnover \citep{supplement}. Therefore, the travelling waves seen at $\R=0$ in the reduced model [Fig.~\ref{fig:Kybif}(a)] are an artifact of the low-order Galerkin truncation. 

We note that the phase-diagram of the full model does not change qualitatively on varying $\Srm$ and $\mu$ \citep{supplement}. Travelling waves exist even for $S < 1$ (the reduced model assumes $\Srm \gg 1$) and emerge at smaller values of $\Prm$ when $\mu$ is reduced.

\begin{figure}
\includegraphics[width=0.85\linewidth]{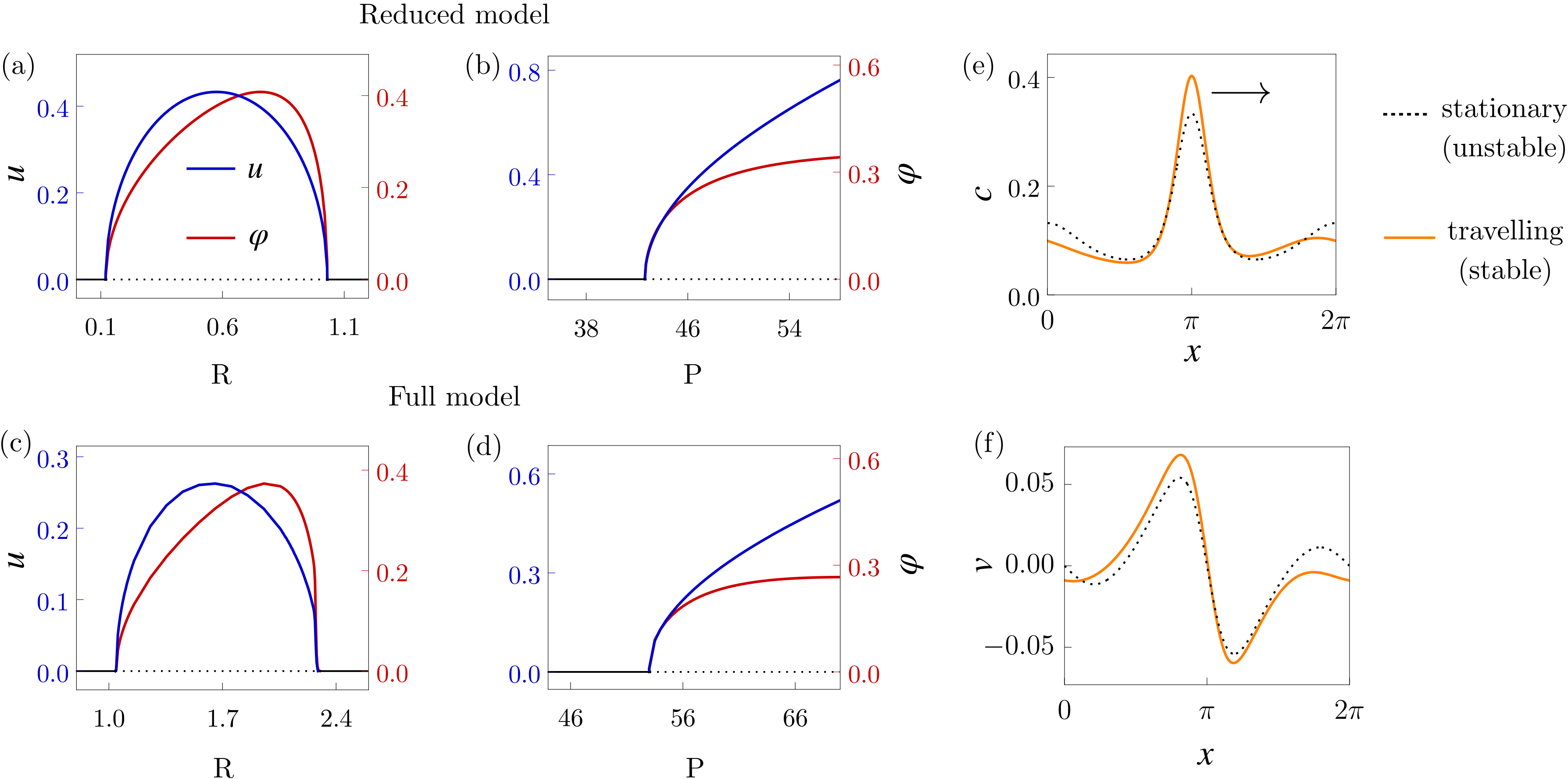}
\caption{Properties of active travelling waves. (a) - (d): Variation of the travelling wave speed $u$ and phase $\varphi$ with $\R$ and $\Prm$; the predictions of the reduced-model are compared with the results of the full model. (e)-(f) Concentration and velocity profiles of the  stable asymmetric travelling wave (solid-orange) and the unstable symmetric stationary pattern (dashed-black), obtained from the full model. Here $\Srm = 5$.}
\label{fig:bifurcation}
\end{figure}

\textbf{Properties of the waves.} To gain insight into the nature of the travelling waves, we return to the reduced model and examine how $\varphi$ and $u$ vary with $\Prm$ and $\R$. Figure~\ref{fig:bifurcation}(a) shows that both $u$ and $\varphi$ vary non-monotonically with $\R$, vanishing at the two bifurcation points and attaining a maximum at an intermediate value of $\R$. Figure~\ref{fig:bifurcation}(b) shows that the speed $u$ increases continuously ($u \sim \sqrt{P}$) while the phase-difference $\varphi$ saturates with increasing $\mathrm{P}$. These features are also exhibited by the numerical solutions of the full model [Fig.~\ref{fig:bifurcation}(c-d)].

A constant $\varphi \ne 0$ implies that the travelling wave is \textit{not} centrosymmetric. This is clearly seen in Fig.~\ref{fig:bifurcation}(e-f), where we compare the concentration and velocity profiles of a stable travelling wave with that of the unstable stationary pattern (computed using numerical continuation \citep{supplement}) that exists at the same parameter values. The asymmetries in the trailing and leading edges of the concentration peak drive a net flux $ j = v c - \partial_x c$ in the direction of propagation. Note that waves travelling in the positive (negative) $x$-direction have positive (negative) values of $u$ and $\varphi$; reflecting a forward-propagating wave yields an equally admissible backward-propagating wave with the same magnitudes of $u$ and $\varphi$ as its counterpart.

The asymmetry of the travelling wave, set by the magnitude of $\varphi$, is clearly controlled by $\R$ [see Figs.~\ref{fig:bifurcation}(a,c)]. Increasing $\Prm$ beyond the onset of waves does not affect their shape but only increases their speed $u$ [see Figs.~\ref{fig:bifurcation}(b,d)]. To understand why a phase-difference is induced only for intermediate values of $\R$, note that the stationary fixed point in the reduced model loses stability along the $\varphi$ subspace when $\rho_2/\rho_1 = \sqrt{ \mathcal{A}_{\varphi}/\mathcal{B}_{\varphi}}\sim \mathcal{O}(1)$ \citep{supplement}. 
In other words, as long as the pattern is dominated by a single spatial mode it remains stationary. The emergence of travelling waves requires two spatial modes of comparable magnitude to interact nonlinearly; this occurs only for intermediate values of $\R$. Our numerical results show that $\rho_2/\rho_1 \approx 1$ at the phase boundary between stationary patterns and travelling waves \citep{supplement}.

\begin{figure*}
\includegraphics[width=\linewidth]{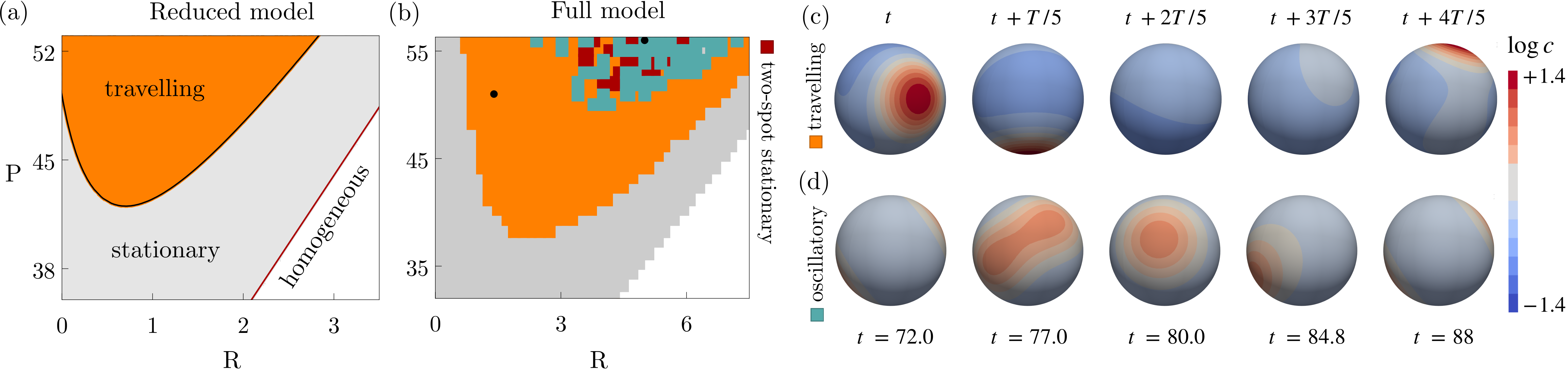}
\caption{Phase-diagram on the surface of a sphere, obtained from (a) the reduced model, and (b) the full model. Snapshots visualizing the evolution of the concentration field obtained from the full model are presented in panels (c) and (d), along with movies in the SI \cite{supplement}. Here $\mu=1$, and $\Srm=4$ (reduced model) and 2 (full model).}
\label{fig:2D_sphere}
\end{figure*}

\textbf{Waves on a sphere.} Since our model is motivated by the actomyosin cortex of cells, we now study a spherical surface and construct a Galerkin-truncated model using scalar [$Y_{lm}(\theta,\phi)$] and vector [$\boldsymbol{\psi}_{lm}(\theta,\phi)$ and $\boldsymbol{\phi}_{lm}(\theta,\phi)$] spherical harmonics, where $\theta$ and $\phi$ are the polar and azimuthal angles respectively \citep{supplement}. In analogy to the one-dimensional periodic domain, we consider  $c= \rho_1 \, e^{-i\varphi_1} \, Y_{1,1}
+ \rho_2 \, e^{-2i\varphi_2} \, Y_{2,2} + c.c. = \sqrt{{3}/{8\pi}}\sin\theta [-2 \rho_{1} \cos(\phi - \varphi_{1})+ \sqrt{5}\rho_{2} \sin\theta \cos(2 [\phi - \varphi_{2}])]$ where $c.c.$ denotes complex conjugation. We find that the equations for the amplitudes $\dot\rho_i$ and phase difference $\dot\varphi = 2(\dot\varphi_2-\dot\varphi_1)$ are exactly the same as \eqref{eq:rho1_pbc}-\eqref{eq:phi_pbc}, but with the coefficients $m_1$, $\mathcal{A}_1$ $\ldots$ listed in the second row of Tab.~\ref{tab:coefficients}.
As such, we obtain the phase diagram for the reduced model shown in Fig.~\ref{fig:2D_sphere}(a). Numerically solving the full model on the sphere using SHTns \citep{schaefferEfficientSphericalHarmonic2013}, we obtain the phase diagram in Fig.~\ref{fig:2D_sphere}(b). In this case as well, the results of the reduced model agree qualitatively with that of the full model. The snapshots in Fig.~\ref{fig:2D_sphere}(c) visualize a typical travelling wave, which manifests as an asymmetric spot translating along an equator of the sphere (see \cite{supplement} for movies). At high $\Prm$, we find oscillatory patterns that involve repeated nucleation and merger of two concentration spots [Fig~\ref{fig:2D_sphere}(d)]. The corresponding region of the $\Prm-\R$ phase diagram appears to be multistable and some initial conditions yield a stationary pattern with two diametrically opposite spots.

\textbf{Parity-breaking and non-reciprocity.} We have demonstrated the emergence of travelling waves on both the line and the sphere when $R \neq 0$. It is important to note that these waves do not result from a Hopf bifurcation of the homogeneous state. Rather, the nonlinear interaction between the inhomogeneous modes of the concentration field generates a non-zero inter-mode phase difference,  leading to a non-centrosymmetric concentration profile that travels with a constant speed. Symmetry analysis of one-dimensional unbounded patterns reveals that such a parity-breaking bifurcation is one of the possible routes to time-dependent states \citep{coulletInstabilitiesOnedimensionalCellular1990}. Indeed, if $c = g(x-\varphi)$ is the symmetric pattern that is subjected to a small antisymmetric perturbation $A\,h(x-\varphi)$ [with $g(-z)=g(z)$ and $h(-z)=-h(z)$], then the weakly nonlinear equations for the amplitude $A$ and phase $\varphi$ take on the scaled form $d_t A = \omega A - A^3$, $d_t \varphi = A$ \citep{coulletInstabilitiesOnedimensionalCellular1990,flessellesDynamicsOnedimensionalInterfaces1991}. Thus, when $\omega>0$, the growth of the parity-breaking perturbation is accompanied by a translation of the entire pattern, just as we see here. Though rather uncommon, this type of bifurcation has previously been observed in the dynamics of interfaces in solidification \citep{Rappel91} and coating flows \citep{Rabaud91}. Such bifurcations typically require nonreciprocity between the interacting modes \cite{fruchartNonreciprocalPhaseTransitions2021}. In our reduced model, the amplitudes $\rho_i$ interact non-reciprocally: the quadratic nonlinearity in \eqref{eq:rho2_pbc} is $\rho_1^2 \cos \varphi$ whereas that in \eqref{eq:rho1_pbc} is $\rho_1 \rho_2 \cos \varphi$.  Our work provides an explicit example of a parity-breaking bifurcation arising from non-reciprocal interactions and leading to the emergence of travelling waves in a simple model of active patterns.

\textbf{Concluding remarks.} In summary, we have shown that a single active species can drive pulsatory patterns in an isotropic fluid film, provided the rate of cytoplasmic exchange is neither too small nor too large.  This holds true even on domains that preclude travelling waves, as seen in simulations on a one-dimensional domain with impermeable boundaries~\citep{supplement}. The corresponding phase diagram is analogous to that in the periodic domain. We have also studied this model on two-dimensional planar domains, with either periodic or no-flux boundaries, and find similar patterns. It is remarkable to note that the model equations \eqref{eqn:conc} and \eqref{eqn:mom} with minimal nonlinear interactions (the advection term and the active stress regulation function) are sufficient to generate spatiotemporal waves on various domains. Furthermore, we find that modifying the model to include cytoplasmic flows or to conserve the total concentration (cortical plus cytoplasmic) alters neither the existence of time-dependent patterns nor the underlying mechanism. Detailed analyses of such extensions are possible directions for future studies. 

The pulsatory patterns seen in our model can be controlled by tuning the turnover rate. N\"aively, one would expect this linear turnover reaction to be simply stabilizing. This is indeed true at the onset of stationary patterns. However, in the strongly nonlinear regime, varying the turnover rate leads to non-reciprocity driven pulsatory patterns. Therefore modulation of the cytoplasmic exchange rate, in response to biological signals, could switch waves on and off. The mechanism elucidated in this study could possibly account for actin waves arising from the interplay between Arp2/3 and formins \cite{chuaCompetitionSynergyArp22024}, Rho-kinase waves seen in extracts of \emph{Xenopus} embryos \cite{landinoRhoFactinSelforganize2021}, as well as waves seen in reconstituted actomyosin networks \cite{krishnaSizedependentTransitionSteady2024}.

\begin{acknowledgments}

\textbf{Acknowledgments}. We acknowledge support of the Department of Atomic Energy, Government of India  (K.V.K. project no. RTI4001) and the Indo–French Centre for the Promotion of Advanced Scientific Research (IFCPAR / CEFIPRA) (J.R.P. project no. 6704-A). J.R.P. acknowledges his Associateship with the International Centre for Theoretical Sciences (ICTS), Tata Institute of Fundamental Research, Bangalore, India. K.V.K. acknowledges the financial support of the John Templeton Foundation (\#62220). The opinions expressed in this paper are those of the authors and not those of the John Templeton Foundation. We thank Vishal Vasan and Aditya Singh Rajput for many useful discussions.
\end{acknowledgments}

\bibliography{references}

\end{document}